\documentclass{ws-procs9x6-cpt19}
\begin{document}

\newcommand{\refeq}[1]{(\ref{#1})}
\def\etal {{\it et al.}}

\title{Establishing a relativistic model for atomic gravimeters}

\author{Ya-jie Wang, Yu-Jie Tan, and Cheng-Gang Shao}


\address{ MOE Key Laboratory of Fundamental Physical Quantities Measurements ${\rm{\& }}$ Hubei Key Laboratory
of Gravitation and Quantum Physics, \\ PGMF and School of Physics, Huazhong University of Science
and Technology, \\ Wuhan 430074, People¡¯s Republic of China}



\begin{abstract}
This work establishes a high-precision relativistic theoretical model: start from studying finite speed of light effect based on a coordinate transformation, and further extend the research methods to analyze the overall relativistic effects. This model promotes the development of testing General Relativity with atomic interferometry.
\end{abstract}

\bodymatter

\section{Introduction}

Since a high-precision gravimeter helps to accurately build a tide model, determine the geoid, test the General Relativity (GR) and so on, it is very necessary to develop high-precision atomic gravimeters and the corresponding theoretical model. Therefore, in the gravity measurements, except for some Newtonian effects, we should also consider some special and general relativistic effects to establish a high-precision theoretical model \cite{1}. In general, researches for the relativistic effects in atomic gravimeters can be divided into two aspects: the finite speed of light (FSL) effect \cite{2,2a,3}, and the GR effects \cite{4,4a}. Since current researches about them exist some disagreements \cite{2,2a,3,4,4a} or being incomplete \cite{4,4a}, we recalculate these effects, and derive a more complete and general expression for them.

\section{Calculation idea for the interferometric phase shift}

According to the working principle of atomic gravimeters, the three-Raman-pulse sequence is usually used to interact with the moving atoms. These pulses split, reflect and recombine the atomic wave packets, respectively. As the evolutions of atoms are space separated, the interferometric signal carries the information of the gravitational field. Thus, the gravitational acceleration can be derived from the measured interferometric phase shift.

A complete atomic interferometry system consists of two parts: atoms and laser lights, which are the main contribution sources to the total interferometric phase. Since the phase is a scalar, it does't depend on the selection of coordinate systems, and we can equivalently observe this atom-laser interacting system in different coordinate systems. In the freely falling system attached to the atoms, all the  relativistic effects are reflected in the laser lights, while in the laser-platform coordinate system fixed on the laboratory, all the  relativistic effects are reflected in the atoms. To a large extent, we here apply the latter idea.
\begin{figure}
\includegraphics[width=4in]{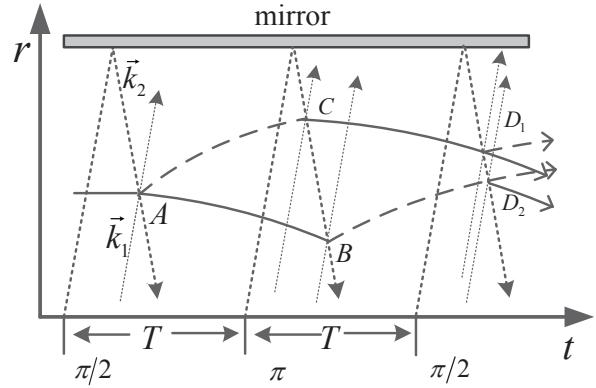}
\caption{A spacetime diagram of a light-pulse atom interferometry. The solid and dashed lines respectively represent the motions of the atoms in the ground and excited states, and the dotted lines stand for the lights manipulating the atoms.}
\label{aba:fig1}
\end{figure}

For a typical atomic gravimeter, one of the two Raman beams is reflected by a mirror (see $\vec{k}_2$ beam in Fig. 1). As $\vec{k}_2$ beam reaches atoms latter than $\vec{k}_1$ beam, and the stimulated Raman transitions occur only when both of the two beams interact with atoms, $\vec{k}_1$ beam can be considered as a ``background light", and the $\vec{k}_2$ beam can be considered as a ``control light".  Based on Fig. 1,  the total interferometric phase shift can be calculated, which contains three parts: the atomic propagation phase shift, laser phase shift, and the separation phase shift. Conventionally, to calculate the relativistic phase shift, one should first solve the geodesic equations of atoms and photons in the general relativistic frame to derive the trajectories of them, solve the five intersections $A$, $B$, $C$, $D_1$, $D_2 $, and then calculate the propagation phase shift with the path integral method, and finally obtain the total interferometric phase shift as well as the gravitational acceleration. However, as the integral intervals for the two paths are temporally different due to the FSL, the calculation is not easy, and usually uses a computer to give a scalar expression. The main idea of this work is to make a coordinate transformation to transfer FSL effect of the ``control light" into the atomic Lagrangian, based on which the integral intervals of the classical Lagrangians for the two paths are temporally same in new coordinate system. The velocity of $\vec{k}_2 $ light becomes $dr'/dt' = \infty $, and the  atomic Lagrangian contains the FSL disturbance in the new system. Then, the total phase shift can be calculated in Bord$\acute{}e$ ABCD matrix and perturbation methods.

\section{Results}
In the special relativistic frame, we calculated the FSL effect, and derived a more complete vectorial expression \cite{5,6}. We find that, except for the results given by Kasevich's group \cite{4,4a} and Steven Chu's group \cite{2a,2}, the FSL correction also includes ${ - 2\frac{{\vec v(T) \cdot {{\vec e}_k}}}{c}\frac{{{\alpha _1} - {\alpha _2}}}{{{{\vec k}_{{\rm{eff}}}} \cdot {{\vec g}_0}}}}$, which has been missed before but at the same magnitude order with the other terms. We further make clear the physical roots of these corrective terms. The main interferometric phase shift arises from the atomic absorbing or emitting laser phase shifts, which can be simply described by
\begin{eqnarray}\label{equation2}
{\phi _{{\rm{laser}}}} = {\vec k_{{\rm{eff}}}} \cdot \vec r - {\omega _{{\rm{eff}}}}t+\phi_0 & \to \left( {\frac{{{\omega _1}{{\vec n}_1} - {\omega _2}{{\vec n}_2}}}{c} + \frac{{{\alpha _1}{{\vec n}_1} - {\alpha _2}{{\vec n}_2}}}{c}T} \right) \cdot \vec r(T,\delta T) \nonumber\\&- \left[ {{\omega _1} - {\omega _2} + ({\alpha _1} - {\alpha _2})T} \right]t(T,\delta T)+\phi_0.
\end{eqnarray}
Here, $\delta T$ is the time delay due to finite propagating speed of light, and $\alpha _1$, $\alpha _2$ are the frequency chirps, which should be introduced to compensate the doppler shift due to atomic motion. We defined the $1/c$ terms related as the FSL effect. Since $\delta T$ is $1/c$ related, the FSL effect includes three parts: the pure FSL time delay, the coupling of the frequency chirp and the time delay, and the chirp-dependence changes of the wave vector. In addition, we find the FSL correction depends on the propagating directions of the lights involved in the measurement process. Therefore, the subterms of FSL correction may be experimentally tested by adjusting the experimental configuration. That's why Cheng et al. \cite{3} reported they only experimentally  verified the FSL effect associated with the coupling of the frequency chirp and the time delay.

Based on the calculation for FSL correction, and derived a more complete relativistic expression of the interferometric phase shift, which is suitable to analyze the atoms moving in three dimensions. In addition, this result first considered the relativistic effects related to Earth's rotation in atomic gravimeters, and also completed the effects related to gravity gradient.

\section{conclusion and prospect}
We mainly developed an analytical study method, based on which the FSL effect is clearly studied, and further a more complete relativistic model for atom gravimeters is established. This work will help to test GR with atomic interferometry. In the near future, on one hand, we will consider exploring the error-elimination schemes, such as the frequency-shift gravity-gradient compensation technique \cite{8}, and in fact we have started this related work \cite{9}; on the other hand, we want to explore the GR-test scheme, such as the test of Lorentz violation and gravitational wave.

\section*{Acknowledgments}
This work is supported by the Post-doctoral Science Foundation of China (Grant Nos. 2017M620308 and 2018T110750).






\begin{thebibliography}{xx}





\bibitem{1}S.\ Wajima \etal, Phys.\ Rev.\ D\ {\bf 55}, 1997 (1964).
\bibitem{2a}A.\ Peters, Ph.D.\ thesis, Stanford University\ (1998).
\bibitem{2}A.\ Peters, K.Y.\ Chung, and S.\ Chu, Metrologia\ {\bf 38}, 25 (2001).
\bibitem{3}B.\ Cheng, P.\ Gillot, S.\ Merlet, and F.\ Pereira Dos Santos, Phys.\ Rev.\ A\ {\bf 92}, 063617 (2015).
\bibitem{4}S.\ Dimopoulos, P.W.\ Graham, J.M.\ Hogan, and M.A.\ Kasevich, Phys.\ Rev.\ Lett.\ {\bf 98}, 111102 (2007).
\bibitem{4a}S.\ Dimopoulos, P.W.\ Graham, J. M.\ Hogan, and M.A.\ Kasevich, Phys.\ Rev.\ D\ {\bf 78}, 042003 (2008).
\bibitem{5}Y.J.\ Tan, C.G.\ Shao, and Z.K.\ Hu, Phys.\ Rev.\ A\ {\bf 94}, 013612 (2016).
\bibitem{6}Y.J.\ Tan, C.G.\ Shao, and Z.K.\ Hu, Phys.\ Rev.\ A\ {\bf 96}, 023604 (2017).
\bibitem{7}Y.J.\ Tan, C.G.\ Shao, and Z.K.\ Hu, Phys.\ Rev.\ D\ {\bf 95}, 024002 (2017).
\bibitem{8}A.\ Roura, Phys.\ Rev.\ Lett.\ {\bf 118}, 160401 (2017)
\bibitem{9}Y.J.\ Wang, X.Y.\ Lu, Y.J.\ Tan, C.G.\ Shao, and Z.K.\ Hu, Phys.\ Rev.\ A\ {\bf 98}, 053604 (2018).





\end{thebibliography}
\end{document}